\documentclass[aps, twocolumn, showpacs]{revtex4-1}
\usepackage{amsmath}
\begin{document}
\title{Comment on'Attempts to test an alternative electrodynamic theory of superconductors by low-temperature scanning tunneling and atomic force microscopy'}
\author{S. C. Tiwari \\
Department of Physics, Institute of Science, Banaras Hindu University, Varanasi 221005, and \\ Institute of Natural Philosophy \\
Varanasi India\\}
\begin{abstract}
It is argued that alternative electrodynamics of superconductivity proposed by Hirsch lacks mathematical rigour and it is conceptually flawed. Gauge non-invariance of the theory makes justification of the experiment reported in [arXiv:1607.05060] to test its prediction doubtful. It seems even with improved techniques the outcome would remain inconclusive.
\end{abstract}
\maketitle

Recent experiment \cite{1} measures the effect of an electrostatic field applied to a niobium superconducting sample using a sharp metal tip in an atomic force microscopy setup. The authors have carefully and critically examined the empirical data to conclude that within the accuracy of their measurements the validity of an alternative superconductivity electrodynamics proposed by Hirsch \cite{2} remains undecided. A natural question arises whether the inconclusiveness of the experimental test is due to the technical limitations of the experiment or the alternative theory being tested has pitfalls/ambiguities. My aim in the present note is to address this question.

Peronio and Giessibl \cite{1} term a London superconductor obeying following equations
\begin{equation}
\frac{\partial{\bf J}}{\partial t} =\frac{n_s e^2}{m} {\bf E}
\end{equation}
\begin{equation}
{\bf \nabla} \times {\bf J} = -\frac{n_s e^2}{mc} {\bf B}
\end{equation}
Here the current density of superconducting electrons is ${\bf J}$, $n_s$ is their number density and m is the free electron mass. The alternative suggested by Hirsch replaces Eq.(1) by
\begin{equation}
\frac{\partial{\bf J}}{\partial t} =\frac{n_s e^2}{m} ({\bf E}+{\bf \nabla} \Phi)
\end{equation}
while retaining Eq.(2) as such. One of the consequences of the modification (3) is that the applied electric field is screened over a London penetration length $\lambda_L$ that arises for the screening of the magnetic fields via Eq.(2). In contrast to this, for a London superconductor the screening length is the Thomas-Fermi screening length $\lambda_{TF}$ just like for a normal metal. The experiment \cite{1} tests this prediction.

To put the matter in perspective recall that Hirsch has raised some interesting and important questions on the BCS theory of superconductivity \cite{3}. The unconventional alternative advocated by him is the theory of hole superconductivity being developed by him and his collaborators for past many years. The perusal of some of his papers related with the topic of the present comment shows that his main objective is to gain support to the hole superconductivity. Koyama \cite{4} has, in fact, offered a nice critique to the issue of the electric field screening length in a superconductor questioning the validity of the assertions made by Hirsch \cite{2,5}. Hirsch in his response \cite{6} notes that the validity of 'the new electrodynamic equations for superconductors' should be decided by experiment. Seen in this light the work reported in \cite{1} would seem to be well motivated. However the theoretical arguments put forward by Hirsch \cite{6} essentially amount to the criticism of BCS theory. On the other hand the set of Equations (1) to (3) is independent of microscopic origin, for example, BCS theory of superconductivity. Therefore, it should be possible to study the validity of Hirsch electrodynamics irrespective of the validity or otherwise of the BCS theory.

An immediate objection to Eq.(3) could be raised: the explicit appearance of the scalar potential $\Phi$ violates the principle of gauge invariance. Since the electromagnetic potentials in classical theory are believed to have no observable physical consequences the theory given by Eqns. (2) and (3) would be untestable in this respect. It may be pointed out that the observed Meissner effect is explained by Eq.(2). The gauge invariance in classical electrodynamics is that only electric and magnetic fields determine the physical quantities and the dynamical equations, whereas the potentials play the role of calculational tools. The arbitrariness in the specification of the potentials is embodied in the gauge transformations
\begin{equation}
{\bf A} ~ \rightarrow ~ {\bf A} + {\bf \nabla}\chi
\end{equation}
\begin{equation}
\Phi ~ \rightarrow ~ \Phi -\frac{1}{c} \frac{\partial \chi}{\partial t}
\end{equation}
such that the electric and magnetic fields are invariant under this transformation
\begin{equation}
{\bf E} = -{\bf \nabla} \Phi - \frac{\partial {\bf A}}{\partial t}
\end{equation}
\begin{equation}
{\bf B} = {\bf \nabla} \times {\bf A}
\end{equation}
In quantum theory the well known Aharonov-Bohm effect \cite{7} ascribes independent physical reality to the potentials in nontrivial physical situations. The magnetic flux quantization in a superconductor ring is an analogous topological effect. The electric field screening implied by Eq.(3) is neither a quantum effect nor a topological one, therefore, here one cannot dispense with the conventional argument denying reality to the potentials. Of course, one could postulate fundamental physical reality to the potentials and develop a new theory. The theory discussed in \cite{2} is not of this kind; it is a modification to the conventional theory and must respect the principle of gauge invariance.

The derivation of Eq.(3) in \cite{2} lacks mathematical rigour. Substituting (7) in Eq.(2)
\begin{equation}
{\bf \nabla} \times {\bf J} = -\frac{n_s e^2}{mc} {\bf \nabla} \times {\bf A}
\end{equation}
Hirsch argues that Eq.(8) can be written as
\begin{equation}
{\bf J} = -\frac{n_s e^2}{mc} {\bf A} 
\end{equation}
Now taking the time-derivative in (9) and using the definition (6) one gets Eq.(3).

What is wrong with this derivation since expression (9) obviously satisfies Eq.(8)? The crucial point is that proceeding from (8) to (9) is not identical with that from (9) to (8). The arbitrariness involved in the curl of a vector field demands that Eq.(8) is equivalent to
\begin{equation}
{\bf J} +{\bf \nabla } \eta = -\frac{n_s e^2}{mc} ({\bf A}+{\bf \nabla} \zeta )
\end{equation}
Solution (9) is a restricted mathematically trivial solution, and on physical grounds one has to justify equating current density to a vector potential which has not been done. Eq.(9) is not fixing a gauge, for example, in Coulomb gauge the vector potential saisfies the condition
\begin{equation}
{\bf \nabla}.{\bf A} =0
\end{equation}
and in the Lorentz gauge one has the condition
\begin{equation}
{\bf \nabla}.{\bf A} +\frac{1}{c} \frac{\partial \Phi}{\partial t}=0
\end{equation}
restricting the allowed gauge transformation (4) and (5). In Eq.(9) there is no gauge freedom.

Hirsch \cite{2} states that 'the right-hand side of this equation is not gauge invariant, while the left-hand side is' referring to Eq.(9). The meaning of this statement is not clear: what is the gauge transformation of current density? I think Hirsch has in mind the Maxwell equation
\begin{equation}
\frac{4\pi}{c} {\bf J} ={\bf \nabla} \times {\bf B} -\frac{1}{c} \frac{\partial {\bf E}}{\partial t}
\end{equation}
since the right-hand side of (13) is gauge invariant. However this is a superficial argument in the context of gauge invariance of ${\bf J}$. Seen in another way in a semiclassical approach it can be verified that Eq.(9) is gauge invariant. The probability current density multiplied by electric charge can be interpreted as current density in Schroedinger theory for electron wavefunction. Generalizing the gauge transformation (4) and (5) with a local phase transformation
\begin{equation}
\Psi ~ \rightarrow ~ \Psi exp(\frac{ie\chi}{\hbar c})
\end{equation}
Eq.(9) is gauge invariant using following expression for the current density
\begin{equation}
{\bf J} =\frac{e\hbar}{2im} (\Psi {\bf \nabla} \Psi^* -\Psi^* {\bf \nabla }\Psi)
\end{equation}
Note that the assumption that $\Psi \Psi^*$ could be  interpreted as number density $n_s$ is not a simple matter in the theory of superconductivity.  Thus the intermediate step (9) makes Hirsch derivation unsound both mathematically and physically.

An alternative to Eq.(1) can be obtained without introducing potentials. Once again the difference in going from Eq.(1) to Eq.(2) or from Eq.(2) to Eq.(1) is important. Hirsch rightly notes that Eq.(2) can be obtained from Eq.(1). Suppose Eq.(2) is postulated then taking its time-derivative one obtains
\begin{equation}
{\bf \nabla } \times \frac{\partial {\bf J}}{\partial t} =-\frac{n_s e^2}{mc} \frac{\partial {\bf B}}{\partial t}
\end{equation}
Using one of the Maxwell equations for $\frac{\partial {\bf B}}{\partial t}$ one gets
\begin{equation}
{\bf \nabla } \times \frac{\partial {\bf J}}{\partial t} =\frac{n_s e^2}{m} {\bf \nabla } \times {\bf E}
\end{equation}
A possible solution is
\begin{equation}
\frac{\partial {\bf J}}{\partial t} +c^2 {\bf \nabla} \rho =\frac{n_s e^2}{m} {\bf E}
\end{equation}
setting arbitrariness in the electric field to be zero. Eq.(18) replaces Eq.(1). In this case there is no gauge invariance issue. Taking divergence of (18) and using the current continuity equation and ${\bf \nabla}.{\bf E} =4\pi \rho$ one gets
\begin{equation}
\nabla^2 \rho -\frac{1}{c^2} \frac{\partial^2 \rho}{\partial t^2} = 4\pi \frac{n_s e^2}{mc^2} \rho
\end{equation}
Eq.(19) differs from Eq.(40d) in \cite{2}, however for vanishing assumed constant charge density $\rho_0$ both become identical.

In conclusion I make two remarks. First, mathematically unsound and conceptually tenuous approach in Hirsch alternative superconductivity electrodynamics is pointed out. Gauge non-invariance of the modified London equation implies that the experimental test reported in \cite{1} cannot be improved by technical precision. It is further suggested that a thorough scrutiny of Hirsch interesting ideas is imperative to devise fruitful experimental tests. Secondly, in my opinion, radically new ideas to develop alternative superconductivity electrodynamics could be explored. In the light of the significance of gauge symmetry it would be interesting to revisit original Weyl gauge theory; a new electrodynamics in the generalized Weyl-Dirac theory \cite{8} could be one such possibility, however further discussion on this is beyond the scope of the present Comment.

{\bf ADDITIONAL NOTE}

Hirsch in the abstract of his paper \cite{2} regarding alternative equations (2) and (3) states that 'These equations resemble equations originally proposed by London brothers but later discarded by them'. It becomes very important to study the original work; fortunately references \cite{9, 10} are accessible. A careful reading of \cite{9} shows that the paper is written with great clarity. Equation (7) in this paper is interpreted in terms of charge density and consequently what the authors term 'independent physical statement' is given by their equation (10) reproduced below
\begin{equation}
\Lambda (\dot{{\bf J}}+ c^2 {\bf \nabla} \rho) ={\bf E}
\end{equation}
The charge density determines electric field as usual
\begin{equation}
{\bf \nabla}.{\bf E} =\rho
\end{equation}
Here $\Lambda =\frac{m}{n e^2} =\frac{4\pi \lambda_L^2}{c^2}$. Obviously in this form Eq.(20) the gauge invariance is a nonissue, and the objections raised in the present comment do not arise. It is noteworthy that the experimental test reported by London \cite{10} is based on this equation. 

Proceeding further, the potentials are introduced assumed to be proportional to the current and the charge densities via Eq.(16) in \cite{9} rendering Eq.(20) in the form that is exactly the Hirsch equation (3). Curiously London does not mention this version in the experimental test \cite{10}. A recent discussion on this experiment \cite{11} seems to ignore this subtle aspect. Relations (16) in \cite{9} cannot be justified in generality though the authors themselves admit that they contain no dynamics. In gauge field theories  one finds an interesting result. Consider abelian Higgs model. The nozero vacuum expectation value of the scalar field leads to the vacuum current proportional to $A^\mu$.

Establishing current continuity based on the Lorentz gauge condition is a superficial argument. In fact, formal similarity of the antisymmetric tensor defined by Eq.(8) in their paper and the definition of the electromagnetic field tensor
\begin{equation}
F_{\mu\nu} =\partial_\mu A_\nu -\partial_\nu A_\mu
\end{equation}
is not sufficient to equate current density 4-vector with the potential 4-vector unless deeper physical implications on the interpretation of electromagnetism as a fluid dynamics of some sort are envisaged \cite{12}.

I am grateful to Dr. A. Peronio for reference [11].

\end{document}